\documentclass[aps,onecolumn,showpacs,showkeys,nofootinbib]{revtex4}
\usepackage[a4paper,top=2cm,bottom=2cm,left=2cm,right=2cm]{geometry}
\usepackage{epsfig}
\usepackage{amsmath}
\usepackage{amsfonts}
\usepackage{amssymb}
\usepackage{graphicx}
\usepackage{colordvi}
\usepackage{subfigure}
\newcommand{\be}{\begin{equation}}
\newcommand{\ee}{\end{equation}}

\begin{document}

\title{A\ two-particle simulation of Nonunitary Newtonian Gravity}
\author{Giovanni Scelza\footnote{%
e-mail address: \textit{lucasce73@gmail.com}}, Filippo Maimone\footnote{%
e-mail address: \textit{filippo.maimone@gmail.com}}, Adele Naddeo$^{(1)}$\footnote{%
e-mail address: \textit{adele.naddeo@na.infn.it}}}
\affiliation{(1)INFN, Sezione di Napoli, C. U. Monte S. Angelo, Via
Cinthia, $80126$ Napoli, Italy}

\begin{abstract}
The result of a numerical simulation of two interacting particles in the
framework of Nonunitary Newtonian Gravity is presented here. Particles are
held close together by a 3-d harmonic trap and interact with each other via
an `electrical' delta-like potential and via the ordinary Newtonian term,
together with a fluctuational nonunitary counterpart of the latter. Fundamental nonunitarity can be seen as arising from the interaction of the physical degrees of freedom with (gravitational) hidden copies of them. Starting
from an energy eigenstate within the ordinary setting, it is shown that,
while energy expectation remains constant, a slow net variation of the von
Neumann entropy for the system as a whole takes place, with a small
modulation induced on the relative entanglement entropy of the two
particles. Besides, the simulation shows explicitly how fundamental
gravity-induced entropy can be clearly distinguished from the subjective
notion of coarse-grained entropy, \textit{i.e.} from the entropy of one
particle with respect to the `environment' of the other.
\end{abstract}

\maketitle
\date{\today }


\section{Introduction}

The possible role of gravity in producing a fundamental nonunitary time
evolution of the state vector in quantum mechanics (QM) has been invoked in
the past few decades by a number of authors, on different grounds \cite%
{0,13,14,10,8}. The long-standing difficulties posed by the two basic
processes of QM, \textit{i.e.} the Schr\"{o}dinger deterministic evolution
and the nonunitary process associated with the act of measurement lead to
the quest of how to reconcile these radically different processes and
explain the transition to classicality.

It is worth noting that the origin of the quantum-to-classical transition is today widely investigated and the emergence of classicality has been attributed mainly to environmental decoherence \cite{zurek1}. A different route to a gravity-related decoherence has been also proposed, which is entirely due to the presence of relativistic time dilation, whose effect on a composite quantum particle is to induce an universal coupling between the center of mass and its internal degrees of freedom. The resulting decoherence of the particle's position takes place even for isolated composite systems and doesn't involve any modification of quantum mechanics \cite{pikovski}. The important point here is that in both the frameworks, envionmental  and time dilation decoherence, global unitarity is preserved and some fundamental problems, at least in our view, remain.

Within this context, in the past years De Filippo introduced a nonunitary
non-Markovian model of Newtonian gravity, NNG from now on (see e.g. Ref.\cite%
{1}, references therein). On the one hand this model can be seen as the
non-relativistic limit of a classically stable version of higher derivative
gravity, whose heuristic application to black hole singularity is in
agreement with Bekenstein-Hawking entropy \cite{bek,bek2,haw}, given the
smoothed singularity of the black hole (see Sec. $VI$ of Ref. \cite%
{1}).
On the other hand, it presents several appealing features to become a
natural candidate as an effective low-energy model of gravity. For example,
while reproducing at a macroscopic level the ordinary Newtonian interaction,
it presents a mass threshold for localization, which for ordinary matter
densities is about $10^{11}$ proton masses \cite{15,noi}. As a key result,
it provides a mechanism for the evolution of macroscopic coherent
superpositions of states into ensembles of pure states. Furthermore the
evolution of the density matrix is compatible with the expectations leading
to the phenomenological spontaneous localization models \cite{10}, as it was
argued that they should be both nonlinear and nonunitary. However, at
variance with them, it does not present obstructions consistent with its
special-relativistic extension \cite{11}.

As to the non-Markovian character of the model, it is interesting to mention
the works of Unruh, who studied some nonunitary toy models mimicking black
hole evaporation, pointing out the quest for non-Markovianity in order to
avoid net dissipation while keeping decoherence \cite{4, unruh}. In the NNG
model such a feature is built-in, together with a number of other appealing
properties, like the natural relativistic extension.

On the black hole entropy side, in the last decades it has been shown how
the decoherence associated to black hole formation and evaporation \cite%
{5,6,7} appears to be a fundamental one, requiring a modification of the
unitary time evolution of quantum mechanics in a nonunitary sense: that
would allow a non vanishing probability for the evolution from pure states
to mixed states. In this sense black hole entropy has been hypothesized \cite%
{ent1} as due to the entanglement between matter-fields inside and outside
the event horizon, the ultimate meaning of the degrees of freedom involved
in its definition remaining quite elusive. Likewise, starting from suitable
initial conditions, only a nonunitary quantum dynamics could allow for a
microscopic derivation of the Second Law of thermodynamics for a closed
system by resorting to the concept of von Neumann entropy, as pointed out in
Ref. \cite{wald}. In this way it should be possible in principle to shed new
light on the long standing and still open problem of quantum foundations of the Second Law of
thermodynamics \cite{th1,th2,th3,th4,th5,th6,th7} without resorting to the
ambiguous and subjective procedure of coarse graining \cite{17,coar}.

Indeed, deep contradictions arise when dealing with the problem of the physical nature and origin of the law of increasing entropy in a closed system, as pointed out by Landau and Lifshitz \cite{th2}. In particular, difficulties show up when one tries to apply statistical physics to the entire Universe taken as a single closed system. A solution could be found by resorting to the general theory of relativity, in which the Universe as a whole must be regarded as a system in a variable gravitational field, but within such a context the application of the law of increase of entropy does not imply relaxation of statistical equilibrium. Furthermore, another problem in formulating the law of increasing entropy arises when defining the direction of time as that of increasing entropy. This should be valid also in quantum mechanics. But quantum mechanics involves a physical non equivalence of the two directions of time, when considering the interaction of a quantum object with a classical one, \textit{i.e.} the measurement process.

Aim of this work is to make a first step in demonstrating the ability of
NNG to reproduce a gravity-induced relaxation towards thermodynamic
equilibrium even for a perfectly isolated system. To do so, a simple system
of two interacting particles is considered. While of course localization
mechanism is not acting in this case, due to the far from threshold mass
of the particles, some of the peculiar characteristic of the model can be
verified in a more general context with respect to the single-lump problem.

More specifically we consider the two particles in an harmonic trap,
interacting with each other through `electrostatic' and gravitational
interaction. Starting from an eigenstate of the physical Hamiltonian, we
carry out a numerical simulation and calculate the time evolution of the von
Neumann entropy. Incidentally, this entropy can be interpreted as an
entanglement entropy with some (fictitious) hidden degrees of freedom, which
recalls the mentioned approach to black hole entropy as entanglement
entropy. As a result, we find that entropy fluctuations take place, owing to
the (nonunitary part of) gravitational interactions, with the initial pure
state evolving into a mixture. This constitutes a first necessary step: an
extension of this work can be envisaged, which could show explicitly within
a realistic setting the way in which thermodynamics could emerge in the
framework of NNG, shedding new light on the possible role of gravity in the
quantum foundations of thermodynamics in ordinary low-energy physics \cite%
{1,15,23,24}.

The plan of the paper is as follows. In Section II we give a brief general
description of the basic NNG model. In Section III we introduce the specific
model which is the subject of our study, and set the parameter values.
Section IV is then devoted to the application of NNG model to the two
particles system via numerical simulations, and contains the main results of
the paper. Finally, in Sec.V, we draw some conclusions and outline future
perspectives of this work. Explicit expressions of the energy matrix
elements are contained in the Appendix.


\section{A brief survey on Nonunitary Newtonian Gravity}

\setcounter{equation}{0} The aim of this Section is to briefly recall the
key features of the NNG model. On the basis of a number of considerations
(among which, the consistency with the basic formal relations of QM, energy
conservation, classical and quantum behavior of matter, suggestion that
non-unitary terms could have a gravitational origin), it is possible to
isolate a two-parameter class of non-unitary gravity models, as discussed in
detail in Refs. \cite{1,15,23,24}. We will comment later in the Section on
these parameters. Let us start here by giving a definition of the model in
its simplest form, suitably oriented to the problem we are going to discuss
below.

Let $\mathcal{H}_{0}[\psi ^{\dagger },\psi ]$ be the non-relativistic
`physical' Hamiltonian of a finite number of particle species, like
electrons, nuclei, ions, atoms and/or molecules, where $\psi ^{\dagger
},\psi $ denote the whole set $\psi _{j}^{\dagger }(\mathbf{x}),\psi _{j}(%
\mathbf{x})$ of creation-annihilation operators, \textit{i.e.} one couple
per particle species and spin component. $\mathcal{H}_{0}[\psi ^{\dagger
},\psi ]$ includes the usual electromagnetic interactions accounted for in
atomic, molecular and condensed-matter physics.

Denoting by $H_{Ph}$ the `physical' energy operator including also the usual
Newtonian interaction,

\bigskip

\begin{equation}
\mathcal{H}_{Ph}[\psi ^{\dagger },\psi ]=\mathcal{H}_{0}[\psi ^{\dagger
},\psi ]-\frac{G}{2}\sum_{j,k}m_{j}m_{k}\int d\mathbf{x}d\mathbf{y}\frac{%
:\psi _{j}^{\dagger }(\mathbf{x})\psi _{j}(\mathbf{x})\psi _{k}^{\dagger }(%
\mathbf{y})\psi _{k}(\mathbf{y}):}{|\mathbf{x-y}|},
\end{equation}

\bigskip

to incorporate that part of gravitational interactions responsible for
non-unitarity one has to introduce complementary creation-annihilation
operators $\widetilde{\psi }_{j}^{\dagger }(\mathbf{x}),\widetilde{\psi }%
_{j}(\mathbf{x})$ and the overall (meta-)Hamiltonian:

\begin{equation}
\begin{split}
\mathcal{H}_{TOT}& =\mathcal{H}_{Ph}[\psi ^{\dagger },\psi ]+\mathcal{H}%
_{Ph}[\tilde{\psi}^{\dagger },\tilde{\psi}] \\
& -\frac{G}{4}\sum_{j,k}m_{j}m_{k}\int d\mathbf{x}d\mathbf{y}\left[ \frac{%
2\psi _{j}^{\dagger }(\mathbf{x})\psi _{j}(\mathbf{x})\tilde{\psi}%
_{k}^{\dagger }(\mathbf{y})\tilde{\psi}_{k}(\mathbf{y})}{|\mathbf{x-y}|}%
\right] \\
& +\frac{G}{4}\sum_{j,k}m_{j}m_{k}\int d\mathbf{x}d\mathbf{y}\left[ \frac{%
:\psi _{j}^{\dagger }(\mathbf{x})\psi _{j}(\mathbf{x})\psi _{k}^{\dagger }(%
\mathbf{y})\psi _{k}(\mathbf{y}):+:\tilde{\psi}_{j}^{\dagger }(\mathbf{x})%
\tilde{\psi}_{j}(\mathbf{x})\tilde{\psi}_{k}^{\dagger }(\mathbf{y})\tilde{%
\psi}_{k}(\mathbf{y}):}{|\mathbf{x-y}|}\right] ,
\end{split}
\label{ge}
\end{equation}

The above operators act on the product $F_{\psi }\otimes F_{\widetilde{\psi }%
}$ of Fock spaces of the $\psi $ and $\widetilde{\psi }$ operators, where $%
m_{i}$ is the mass of the $i$-th particle species and $G$ is the
gravitational constant. The $\widetilde{\psi }$ operators obey the same
statistics as the corresponding operators $\psi $, while $[\psi ,\widetilde{
\psi }]_{-}=[\psi ,\widetilde{\psi }^{\dagger }]_{-}=0$.

The meta-particle state space $S$ is the subspace of $F_{\psi }\otimes F_{
\widetilde{\psi }}$, including the meta-states obtained from the vacuum $%
\left\vert \left\vert 0\right\rangle \right\rangle =\left\vert
0\right\rangle _{\psi }\otimes \left\vert 0\right\rangle _{\widetilde{\psi }%
} $ by applying operators built in terms of the products $\psi _{j}^{\dagger
}( \mathbf{x})\widetilde{\psi }_{j}^{\dagger }(\mathbf{y})$ and symmetrical
with respect to the interchange $\psi ^{\dagger }\leftrightarrow \widetilde{
\psi }^{\dagger }$; as a consequence they have the same number of $\psi $
(physical) and $\widetilde{\psi }$ (hidden) meta-particles of each species.
Since constrained meta-states cannot distinguish between physical and hidden
operators, the observable algebra is identified with the physical operator
algebra. In view of this, expectation values can be evaluated by
preliminarily tracing out the $\widetilde{\psi }$ operators. In particular,
the most general meta-state corresponding to one particle states is
represented by
\begin{equation}
\left\vert \left\vert f\right\rangle \right\rangle =\int d\mathbf{x}\int d%
\mathbf{y}\,f(\mathbf{x},\mathbf{y})\psi _{j}^{\dagger }(\mathbf{x})%
\widetilde{\psi }_{j}^{\dagger }(\mathbf{y})\left\vert \left\vert
0\right\rangle \right\rangle ,  \label{f}
\end{equation}%
with
\begin{equation*}
f(\mathbf{x},\mathbf{y})=f(\mathbf{y},\mathbf{x}).
\end{equation*}%
This is a consistent definition since $\mathcal{H}_{TOT}$\ generates a group
of (unitary) endomorphisms of $S$.

If we prepare a pure $n$-particle state, represented in the original
setting, excluding gravitational interactions, by%
\begin{equation*}
\left\vert g\right\rangle =\int d^{n}\mathbf{x}\,g\left( \mathbf{x}_{1},%
\mathbf{x}_{2},...,\mathbf{x}_{n}\right) \psi _{j_{1}}^{\dagger }(\mathbf{x}%
_{1})\psi _{j_{2}}^{\dagger }(\mathbf{x}_{2})...\psi _{j_{n}}^{\dagger }(%
\mathbf{x}_{n})\left\vert 0\right\rangle ,
\end{equation*}%
its representation in $S$ is given by the meta-state
\begin{equation*}
\int d^{n}\mathbf{x}\ d^{n}\mathbf{y}\biggl(g\left( \mathbf{x}_{1},...,%
\mathbf{x}_{n}\right) g\left( \mathbf{y}_{1},...,\mathbf{y}_{n}\right)
\times \psi _{j_{1}}^{\dagger }(\mathbf{x}_{1})...\psi _{j_{n}}^{\dagger }(%
\mathbf{x}_{n})\ \widetilde{\psi }_{j_{1}}^{\dagger }(\mathbf{y}_{1})...%
\widetilde{\psi }_{j_{n}}^{\dagger }(\mathbf{y}_{n})\left\vert \left\vert
0\right\rangle \right\rangle \biggr).
\end{equation*}

It should be stressed that the model could be formulated, through an
Hubbard-Stratonovich transformation, without any explicit reference to the
hidden degrees of freedom, as a dynamic equation for the density matrix of
the observed degrees of freedom \cite{1}.

Note that $\mathcal{H}_{TOT}$ and $\mathcal{H}_{Ph}[\psi ^{\dagger },\psi ]+%
\mathcal{H}_{Ph}[\tilde{\psi}^{\dagger },\tilde{\psi}]$ differ only due to
correlations. In fact, because of the state constraint, the hidden degrees
of freedom show the same average energy of the observed ones, while the two
last sums in $\mathcal{H}_{TOT}$ have approximately equal expectation values
and fluctuate around the classical gravitational energy. While these energy
fluctuations have to be present in any model leading to dynamical wave
function localization, which in itself requires a certain injection of energy%
\cite{squires1}, on the other side the same fluctuations, though irrelevant
on a macroscopic scale, are precisely what can lead to thermodynamic
equilibrium in a closed system if thermodynamic entropy is identified with
von Neumann entropy\cite{kay}. Indeed, due to the interaction with the
hidden degrees of freedom, a pure eigenstate of the ordinary energy $%
\mathcal{H}_{Ph}$ is expected to evolve into a mixture. It is just this
feature, inferred on the above general grounds, that we will check
numerically on a specific (simple) physical system.

A further comment is in order. Within a more general version of the model,
we have $N$ interacting identical copies of the system under study: indeed
such a symmetry requirement, which has to be imposed as a constraint on
meta-states, is mandatory in order to avoid \textit{a priori} a net energy flux
between observable and hidden degrees of freedom and to guarantee energy
conservation. That happens in strict analogy with the nonunitary toy model
introduced by Unruh and Wald \cite{4} and the simple quantum mechanical
model recently analysed by Unruh \cite{unruh}, which takes into account the
scattering of two particles off each other, mediated by some hidden degrees
of freedom (represented as a certain number of spin $\frac{1}{2}$ objects).
It is also interesting to note that the limit $N\rightarrow \infty $
reproduces the famous non-linear Newton-Schr\"{o}dinger equation, sometimes
considered in the literature as a possible candidate equation for the
inclusion of self-gravity in QM, relevant to the quantum-classical
transition \cite{10,15}.


\section{The physical system: interacting particles in a trap}

\setcounter{equation}{0} In this Section we introduce the system model that
will be the subject of our study in the following. Then we look for a range
of physical parameters, where the two-body interaction term in the
Hamiltonian can be considered at most comparable with the kinetic and
harmonic terms.

Let's start by considering a single particle of mass $\mu $, confined by a
spherically symmetric harmonic trap \cite{dav}. The complete wave-functions
are
\begin{equation}
\psi _{\mathfrak{nlm}}(r,\theta ,\phi )=R_{\mathfrak{nl}}(\xi )Y_{\mathfrak{l%
}}^{\mathfrak{m}}(\theta ,\phi ),  \label{prima}
\end{equation}%
where $\xi =r\sqrt{\frac{\mu \omega }{\hbar }},$ $R_{\mathfrak{nl}}=A_{%
\mathfrak{nl}}\frac{1}{\xi }e^{-\frac{\xi ^{2}}{2}}\xi ^{\mathfrak{l}+1}%
\mathbf{F}(-\mathfrak{n},\mathfrak{l}+\frac{3}{2},\xi ^{2})$ with $A_{%
\mathfrak{nl}}$ the normalization factor and $\mathbf{F}$ the confluent
hypergeometric function and $Y_{\mathfrak{l}}^{\mathfrak{m}}(\theta ,\phi )$
are the spherical harmonics. The corresponding energy levels are
\begin{equation}
E_{\mathfrak{n,l}}=\hbar \omega \biggl(2\mathfrak{n}+\mathfrak{l}+\frac{3}{2}%
\biggr)\qquad \mathfrak{n,l}=0,1,2\dots .  \label{due}
\end{equation}%
Let us now introduce a system of $n-$interacting particles in the same
spherical trap, whose `physical' Hamiltonian, in the ordinary
(first-quantization) setting, is
\begin{equation}
H_{Ph}\left( \mathbf{x}_{1},\mathbf{x}_{2}\right) =-\left( 1/2\right)
\sum\limits_{i=1}^{n}\biggl(\frac{\hbar ^{2}}{\mu }\Delta _{\mathbf{x}%
_{i}}-\mu \omega ^{2}\mathbf{x}_{i}^{2}\biggr)+\frac{4\pi \hbar ^{2}l_{s}}{%
\mu }\sum\limits_{i<j=1}^{n}\delta ^{(3)}(\mathbf{x}_{i}-\mathbf{x}%
_{j})-G\mu ^{2}\sum\limits_{i<j=1}^{n}\left\vert \mathbf{x}_{i}-\mathbf{x}%
_{j}\right\vert ^{-1};  \label{mol1}
\end{equation}%
here $\mu $ is the mass of the particles, $\omega $ is the frequency of the
trap and $l_{s}$ is the s-wave scattering length. We are considering a
`dilute' system, such that the electrical interaction can be assumed to have
a contact form with a dominant s-wave scattering channel. We take numerical
parameters that make the electrical interaction at most comparable
with the oscillator's energy, while gravity enters the problem as an higher
order correction. Reliable values of the parameters could be estimated such
that the condition

\begin{equation}
\eta =\frac{U}{\left( 3/2\right) \hbar \omega }\lesssim 1  \label{ratio1}
\end{equation}%
holds, where $U$ is the mean value of the delta interaction potential
evaluated on the system's initial state. Furthermore, if we assume that only
the first levels above the ground are excited, then U can be evaluated
approximately on the $0-$th order ground state

\begin{eqnarray}
U \backsimeq\frac{4\hbar ^{2}l_{s}}{\mu \sqrt{\pi }}\left( \frac{\mu \omega
}{\hbar } \right) ^{3/2}.  \label{ratio2}
\end{eqnarray}



In our case, then, $\eta=0.98.$\newline
From now on we fix $n=2$ in Eq. (\ref{mol1}), which corresponds to the
simplest case of two interacting particles. In the following Section, we
apply NNG model to this system and study its time evolution, starting from
an initial pure state, particularly from an energy eigenstate. For
simplicity, we perform our computation within a reduced Hilbert space built
on the lowest energy eigenvalues of the system, corresponding to the tensor
product of the single-particle space spanned by the first 4 states, \textit{i.e.}
the ground state and three degenerate excited states of the harmonic
oscillator. The consistency of this huge dimensionality reduction is
verified \textit{a posteriori} by the fact that higher energy levels are not
excited at all throughout the time evolution.

\section{Numerical results - time evolution of the von Neumann entropy}

\setcounter{equation}{0}The aim of this Section is to study the time
evolution of the two-particles system introduced above in the framework of
the NNG model \cite{1,15,23,24} starting from a pure state for each
particle, which we identify with the lowest energy state defined by the
quantum number $\mathfrak{n}=0$ in (\ref{due}). We will define the general
meta-Hamiltonian by coupling via gravitational interaction only the physical
system with an hidden system which is the identical copy of it. Then the
time dependent physical density matrix is computed by tracing out the hidden
degrees of freedom and the corresponding von Neumann entropy is derived as
the entanglement entropy with such hidden degrees of freedom. The nonunitary
physical theory obtained will produce an entropy fluctuation in the closed
system chosen and then a fundamental gravitationally-induced decoherence,
which can be clearly distinguished from the usual coarse-grained entropy. We
carry out numerical simulations by fixing for simplicity a number of
particles $n=2$, which is the simplest nontrivial case one can consider in
order to show some of the most important peculiarities of NNG. Our model, in
the (first-quantization) ordinary setting, is defined by

\begin{equation*}
H_{TOT}=H_{Ph}\left( \mathbf{x}_{1},\mathbf{x}_{2}\right) +H_{Ph}\left(
\widetilde{\mathbf{x}}_{1},\widetilde{\mathbf{x}}_{2}\right) +H_{NNG}\left(
\mathbf{x}_{1},\mathbf{x}_{2};\widetilde{\mathbf{x}}_{1},\widetilde{\mathbf{x%
}}_{2}\right) ,
\end{equation*}

with $H_{NNG}=G\mu^2\sum_{i<j}\biggl(-\frac{1}{\vert \mathbf{x}_i-\widetilde{%
\mathbf{x}}_j\vert}+\frac{1}{2\vert \mathbf{x}_i-\mathbf{x}_j\vert}+\frac{1}{%
2\vert \widetilde{\mathbf{x}}_i-\widetilde{\mathbf{x}}_j\vert}\biggr)$ in
agreement with (\ref{ge}).


We write the components of the generic basis meta-state $\Vert \Psi \rangle
\rangle $ as
\begin{equation}
\Vert \Psi _{\alpha }\rangle\rangle =\left\vert
i_{1}i_{2}...i_{n}\right\rangle \otimes \left\vert
j_{1}j_{2}...j_{n}\right\rangle =\left[ \bigotimes\limits_{a=1}^{n}\left%
\vert i_{a}\right\rangle \right] \otimes \left[ \bigotimes\limits_{b=1}^{n}%
\left\vert j_{b}\right\rangle \right] ,  \label{metastate1}
\end{equation}%
where $\alpha $ defines the following cumulative index $\alpha \equiv \left(
i_{1}i_{2}...i_{n};j_{1}j_{2}...j_{n}\right) $ and $i_{a}=1,2,3,4$ is one of
the four basis states of the single molecule (the quantum numbers we are
dealing with are $\mathfrak{n}=0$; $\mathfrak{l}=0,1$; $\mathfrak{m}=-1,0,1$%
). In the following we carry out numerical simulations for a system of $n=2$
particles, so we get $16$ basis states and a general meta-state with $256$
components respectively, while the matrix dimension of the whole
meta-Hamiltonian $H_{TOT}$ is $64\times 64$. The evaluation of the matrix
elements for the relevant (nonunitary) gravitational interaction between the
basis meta-states (\ref{metastate1}) is given in detail in the Appendix.

Let us now outline the main steps of our simulation. A numerical
diagonalization gives the eigenvalues and the eigenvectors of $H_{TOT}$, so
we get all the ingredients to study the time evolution of the system. By
choosing as initial state the following general unentangled meta-state:
\begin{equation}
\Vert \Psi \left( 0\right) \rangle \rangle =\sum_{I=1}^{4^{2n}}\alpha _{I}\
\Vert \Psi _{I}\rangle \rangle ,  \label{m1}
\end{equation}%
where $\Vert \Psi _{I}\rangle \rangle $ are the meta-eigenstates of $H_{TOT}$%
, the time evolution will produce the entangled meta-state:
\begin{equation}
\Vert \Psi \left( t\right) \rangle \rangle =\sum_{I=1}^{4^{2n}}\alpha
_{I}\exp \left[ -\left( i/\hbar \right) H_{TOT}\ t\right] \Vert \Psi
_{I}\rangle \rangle .  \label{m2}
\end{equation}

\begin{figure}[tbph]
\centering
\includegraphics[scale=1]{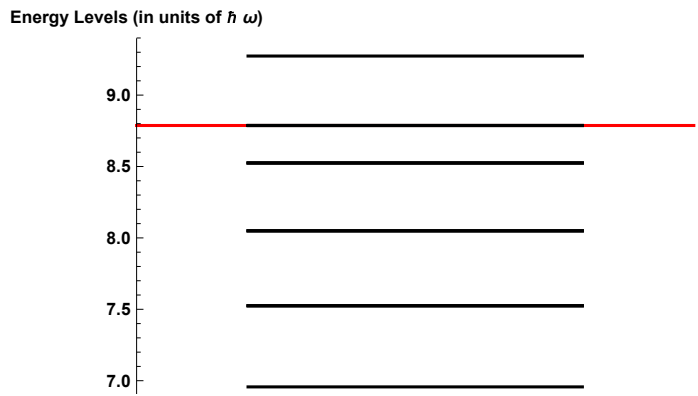}%
\caption{ Energy levels of the $n=2$-particle system. In red the energy
level of the eigenstate $\vert\phi _{2}\rangle$}
\label{fig:fig1}
\end{figure}

\begin{figure}[tbph]
\centering
\includegraphics[scale=1]{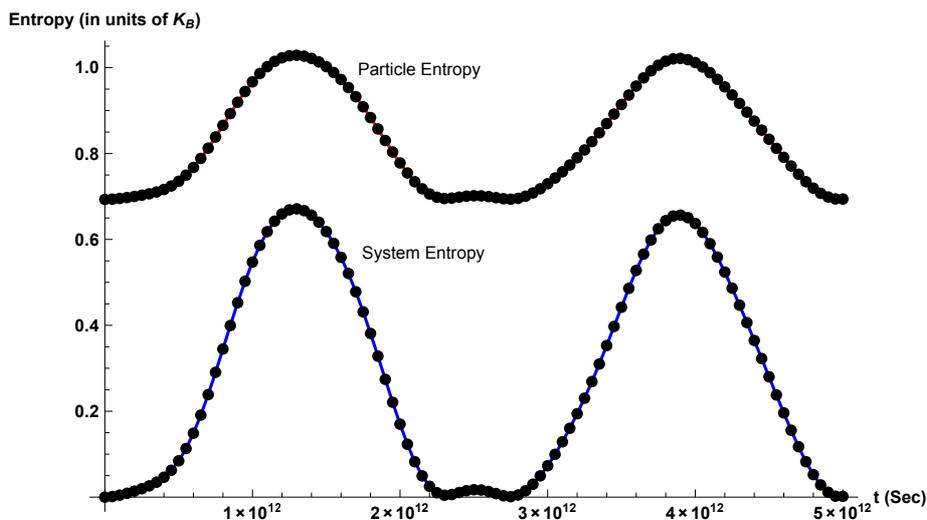}
\caption{ Time evolution of one-particle and two-particle entropies (respectively, curve above and below) for the initial state $\phi _{2}$.}
\label{fig:fig2}
\end{figure}

It is now possible to calculate the density matrix of only one molecule of
the physical system by tracing out the remaining $2n-1$ degrees of freedom:
\begin{equation}
\left\langle i_{1}\right\vert \rho _{m}\left( t\right) \left\vert
j_{1}\right\rangle =\sum\limits_{i_{2},i_{3},...,i_{2n}=1}^{4}\left\langle
\left\{ i_{1},i_{2},...,i_{2n}\right\} \right. \Vert \Psi \left( t\right)
\rangle \rangle \left\langle \left\langle \Psi \left( t\right) \right\vert
\right\vert \left. \left\{ j_{1},i_{2},...,i_{2n}\right\} \right\rangle .
\label{md1}
\end{equation}%
But the key quantity for our work is the density matrix of the physical
system, which is obtained by tracing out the hidden degrees of freedom. That
corresponds to split the whole meta-system into two parts, as every
bipartite system, identified with the physical and hidden degrees of freedom
respectively. We get:
\begin{eqnarray}
&&\left\langle \left\{ i_{1},..,i_{n}\right\} \right\vert \rho _{PH}\left(
t\right) \left\vert \{j_{1},..,j_{n}\}\right\rangle =  \notag \\
&=&\sum\limits_{i_{n+1},...,i_{2n}=1}^{4}\left\langle \left\{
i_{1},i_{2},..,i_{n},i_{n+1},..,i_{2n}\right\} \right. \Vert \Psi \left(
t\right) \rangle \rangle \left\langle \left\langle \Psi \left( t\right)
\right\vert \right\vert \left. \left\{
j_{1},j_{2},..,j_{n},i_{n+1},..,i_{2n}\right\} \right\rangle .  \label{mdph1}
\end{eqnarray}%
Starting from the von Neumann entropy definition $\mathcal{S}(\rho
)=-k_{B}Tr(\rho \log \rho )$, where $k_{B}$ is the Boltzmann constant, we
compute the single-particle von Neumann entropy $S(\rho _{m})$, and the
two-particle system von Neumann entropy $S(\rho _{PH})$. The results shown
in the following refer to an initial state, $\vert\phi _{2}\rangle,$ equal to the eigenstate of the
physical energy  associated with the $2^{nd}$ highest energy
eigenvalue $E_{2}$ shown in fig. \ref{fig:fig1}. The results for the
entropies are shown in fig. \ref{fig:fig2}
respectively for the following values of the physical parameters $l_{s},\mu $
and $\omega $: $l_{s}=5.5\cdot 10^{-8}m$, $\mu =1.2\cdot 10^{-24}Kg$ and $%
\omega =4\pi \cdot 10^{3}s^{-1}$ (which are compatible with current
experiments with trapped ultracold atoms \cite{exp1} and complex molecules
\cite{exp2}) and with an artificially augmented 'gravitational constant' $%
G=6.67408\times 10^{-6}m^{3}kg^{-1}s^{-2}$ ($10^{5}$ times the real
constant).

\bigskip To begin with, due to gravity-induced fluctuations the system
entropy shows a net variation over very long times, at variance with the
case $G=0$, in which it would have been a constant of motion. At the same
time, single particle entropy, which in the ordinary setting is itself
constant, shows now a modulation by the system entropy itself. Incidentally,
it has been verified that the expectation of physical energy is a constant,
meaning that the non-unitary term has no net energy associated with itself,
but is purely fluctuational.

\begin{figure}[tbph]
\centering
\includegraphics[scale=1]{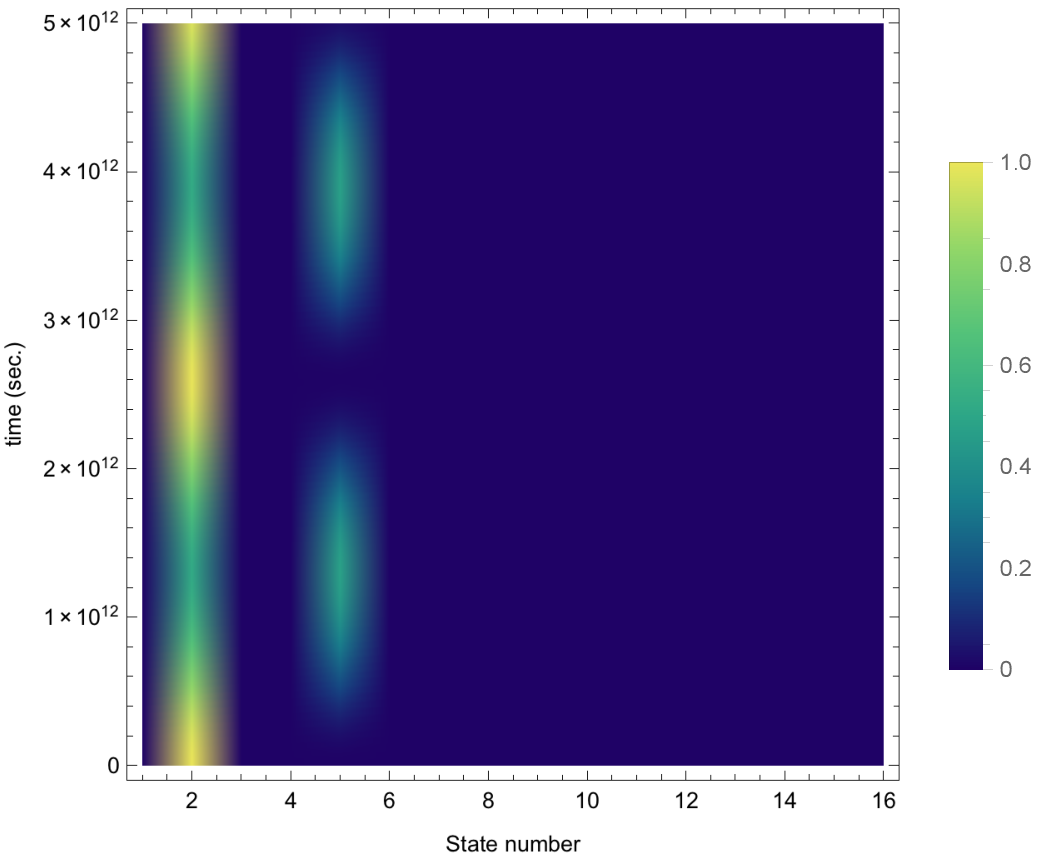}
\caption{Density plot of the energy levels in the case of an initial state
of the system equal to $\vert\phi _{2}\rangle$. It is possible to observe the
involvement of another energy level starting from a time $\sim 10^{12}$
sec. This value is in agreement with the time-energy uncertainty relation $%
\Delta t\sim \frac{\hbar }{\protect\sqrt{\langle H_{Tot}^{2}\rangle -\langle
H_{Tot}\rangle ^{2}}}$ applied in the meta-Hilbert space and evaluated
on the initial meta-state. An estimate in terms of the basic parameters of the problem is given by $\Delta t\thicksim \hslash^{3/2}G^{-1}\mu^{-5/2}\omega^{-1/2}$, obtained by considering the gravitational energy corresponding to the length scale of the fundamental state of the harmonic oscillator.}
\label{fig:fig4}
\end{figure}

Besides the density plot (see fig. \ref{fig:fig4}) shows that another energy level is slightly
excited, leading to the initial entropy growth.


To ensure physical consistency with our assumption that the whole
(meta-)dynamics is confined within the truncated $64$-dimensional Hilbert
space, one can assume suitable anharmonic corrections to the harmonic trap
potential, leading to a sufficient spacing of higher energy levels for these
levels not to be excited at all by all the interactions in the system.

A further comment is in order about the employed value of gravitational
constant throughout the computation. The computation in fact has been carried out keeping
experimentally accessible values of particle masses and other parameters
like scattering length and trap frequency, while using a significantly
higher value of the real gravitational constant. Using the real value for the
last constant would have produced a numerical noise overcoming the small
interesting effect we wanted to see.\newline
Alternatively, one can use the real value of $G$ and far from reachable
values of the other parameters, adopting for example the following scaling of
the Schroedinger equation:

\begin{equation*}
i\hbar \frac{\partial }{\partial t}\Psi (X,Y;t)=H_{TOT}\Psi (X,Y;t),
\end{equation*}%
\begin{equation*}
G\rightarrow \lambda G,\ \ \ \ \ \ t\rightarrow t,\ \ \ \ \ \ \mu
\rightarrow \lambda ^{-2/5}\mu ,\ \ \ \ \ \ \ X\rightarrow \lambda ^{1/5}X,\
\ \ \ \ \ l_{s}\rightarrow \lambda ^{1/5}l_{s}
\end{equation*}%
\begin{equation*}
\Psi (X,Y;t)\rightarrow \Psi (\lambda ^{-1/5}X,\lambda ^{-1/5}Y;t).
\end{equation*}
Incidentally,
starting from a specific numerical solution, the scaling above provides also
for a continuous one-parameter family of solutions.\\

\bigskip

It is worth noting that the simple multiperiodic behavior observed for the entropy is due to the very elementary structure of our system's energy levels, together with the fact that the involved gravitational energy is small compared to the separation between levels. Taking a system with a macroscopic number of particles, gravitational energy would very easily overcome the tiny level spacings of the order of $10^{-n}$ \cite{th2}. As this condition is met, and so many different frequencies enter into play, we can expect that entropy fluctuations would be strongly suppressed within a timescale proportional to $n^{-2}$ (as can be inferred from the time-uncertainty relation).


\section{Conclusions and perspectives}

The main feature introduced by NNG model is a fundamental non-unitarity
arising from the gravitational interaction between the physical and hidden
degrees of freedom, the last being an identical copy of the physical system,
and with a symmetry constraint on the state-space. As a consequence,
taking the simplest nontrivial (almost realistic) physical system of
two-interacting particles the initially pure state taken as an eigenstate of
the physical Hamiltonian (including the ordinary Newtonian potential)
`diffuses' and develops over a very long time in a mixture of the states
involving neighboring energy levels. As a consequence both the system
entropy and the single particle entropy manifest a net variation, while the
expectation of energy remains constant over all the time. The very long time scale of the system studied here is compatible with the up to now experience with very small and isolated  systems like the systems used to implement basic quantum computation, in which quantum coherence is preserved as long as environmental decoherence is controlled.

One may easily expect that using a more complex system with many particles
one would get a much more rapid growth and subsequent stabilization of the
system entropy, following the formation of a microcanonical ensemble around
the initial energy level. In fact the accessible Hilbert space is rapidly
growing as new degrees of freedom are added. The entropy is expected to grow
consequently, while the spacing of energy levels, being proportional to
$exp(-S)$, becomes infinitesimally small. This last circumstance would
allow for the excitation of a huge number of surrounding energy levels,
and transitions among them would be induced even by the very weak
gravitational interactions.

Of course, on the basis of the Poincare's recurrence theorem applied in the
meta-Hilbert space, one may expect that after a sufficiently long time the
system entropy would diminish, going back to zero. But the time at which
this would happen becomes longer and longer as the number of particles gets
thermodynamic relevance.

It is important to stress that the analysis performed above gives also an
(in principle) operational way to clearly distinguish between the ordinary
coarse-graining (subjective) entropy and the 'fundamental' entropy due to
the peculiar non unitary gravitational interactions of the model.

As a future perspective of the present work, a system of thermodynamic
size should be analytically solved and studied in order to ensure a definite
and real ability of the NNG model to reproduce the spontaneous relaxation of
mesoscopic and macroscopic systems toward thermodynamic equilibrium, paving
the way to a new\ foundational basis for the Second Law of thermodynamics.

\section*{Acknowledgements}

One of the authors (G.S) is indebted with Francesco Siano for helpful
support and useful discussions.

\appendix

\section{Gravitational interaction matrix elements}

\setcounter{equation}{0} This Appendix is devoted to the evaluation of the matrix elements of the
gravitational interaction term

\begin{equation}
H_{G}=-G\mu ^{2}\sum_{i<j}\frac{1}{\left\vert \mathbf{x}_{i}-\widetilde{%
\mathbf{x}}_{j}\right\vert }  \label{g1}
\end{equation}
in the basis of meta-states with $m^{2n}$ components, as defined in Eq. (\ref%
{metastate1}), where $m$ is the number of single-particle basis states chosen. 
We have

\begin{eqnarray}
&&\left\langle \left\{ i_{1},i_{2},...,i_{2n}\right\} \left\vert \ H_{G}\
\right\vert \left\{ j_{1},j_{2},...,j_{2n}\right\} \right\rangle  \label{g2}
\\
&=&-G\mu ^{2}\sum_{\alpha =1}^{n}\sum_{\beta =n+1}^{2n}\iint d\mathbf{r}%
_{\alpha }d\mathbf{\widetilde{r}}_{\beta }\psi _{i_{\alpha }}^{\ast }\left(
\mathbf{r}_{\alpha }\right) \psi _{i_{\beta }}^{\ast }\left( \mathbf{%
\widetilde{r}}_{\beta }\right) \frac{1}{|\mathbf{r}_{\alpha }-\mathbf{%
\widetilde{r}}_{\beta }|}\psi _{j_{\alpha }}\left( \mathbf{r}_{\alpha
}\right) \psi _{j_{\beta }}\left( \mathbf{\widetilde{r}}_{\beta }\right) .
\notag
\end{eqnarray}

\bigskip

In order to evaluate the general matrix element let us concentrate on the
integral and introduce spherical coordinates. We get

\begin{eqnarray*}
&&\iint d\mathbf{r}_{1}d\mathbf{r}_{2}\frac{\psi _{i}^{\ast }\left( \mathbf{r%
}_{1}\right) \psi _{j}^{\ast }\left( \mathbf{r}_{2}\right) \psi _{i^{\prime
}}\left( \mathbf{r}_{1}\right) \psi _{j^{\prime }}\left( \mathbf{r}%
_{2}\right) }{\left\vert \mathbf{r}_{1}\mathbf{-r}_{2}\right\vert }%
=\sum_{l=0}^{\infty }\iint_{0}^{\infty }dr_{1}dr_{2}\left( r_{1}r_{2}\right)
^{2}\times \\
&\times &\frac{r_{<}^{l}}{r_{>}^{l+1}}R_{i}^{\ast }\left( r_{1}\right)
R_{j}^{\ast }\left( r_{2}\right) R_{i^{\prime }}\left( r_{1}\right)
R_{j^{\prime }}\left( r_{2}\right) \iint d\Omega _{1}d\Omega
_{2}Y_{l_{i}}^{m_{i}\ast }\left( \theta _{1},\phi _{1}\right)
Y_{l_{j}}^{m_{j}\ast }\left( \theta _{2},\phi _{2}\right) \times \\
&\times &Y_{l_{i^{\prime }}}^{m_{i^{\prime }}}\left( \theta _{1},\phi
_{1}\right) Y_{l_{j^{\prime }}}^{m_{j^{\prime }}}\left( \theta _{2},\phi
_{2}\right) P_{l}\left( \cos \gamma \right) ,
\end{eqnarray*}

\bigskip

where we have used the multipolar expansion $\frac{1}{\left\vert \mathbf{r}%
_{1}\mathbf{-r}_{2}\right\vert }=\sum_{l=0}^{\infty }\frac{r_{<}^{l}}{%
r_{>}^{l+1}}P_{l}(\cos \gamma )$ ($r_{<}$($r_{>}$) are the minor (major)
between $r_{1}$\textbf{(}$r_{2}$) and $\gamma $ is the angle between the
orientations $\left( \theta _{1},\phi _{1}\right) $ and $\left( \theta
_{2},\phi _{2}\right) $). By substituting the expression for $P_{l}(\cos
\gamma )$

\bigskip

\begin{equation}
P_{l}(\cos \gamma )=\frac{4\pi }{2l+1}\sum_{m=-l}^{l}Y_{l}^{m\ast }\left(
\theta _{1},\phi _{1}\right) Y_{l}^{m}\left( \theta _{2},\phi _{2}\right) ,
\label{g4}
\end{equation}

\bigskip

we obtain

\begin{eqnarray}
&&\sum_{l=0}^{1}\frac{4\pi }{2l+1}\iint_{0}^{\infty }dr_{1}dr_{2}\left(
r_{1}r_{2}\right) ^{2}\frac{r_{<}^{l}}{r_{>}^{l+1}}R_{i}^{\ast }\left(
r_{1}\right) R_{j}^{\ast }\left( r_{2}\right) R_{i^{\prime }}\left(
r_{1}\right) R_{j^{\prime }}\left( r_{2}\right) \times  \notag \\
&\times &\sum_{m=-l}^{l}\left\{ \int d\Omega _{1}Y_{l_{i}}^{m_{i}\ast
}\left( \theta _{1},\phi _{1}\right) Y_{l_{i^{\prime }}}^{m_{i^{\prime
}}}\left( \theta _{1},\phi _{1}\right) Y_{l}^{m\ast }\left( \theta _{1},\phi
_{1}\right) \right\} \times  \label{g5} \\
&\times &\left\{ \int d\Omega _{2}Y_{l_{j}}^{m_{j}\ast }\left( \theta
_{2},\phi _{2}\right) Y_{l_{j^{\prime }}}^{m_{j^{\prime }}}\left( \theta
_{2},\phi _{2}\right) Y_{l}^{m}\left( \theta _{2},\phi _{2}\right) \right\} .
\notag
\end{eqnarray}%
The two angular integrations can be performed by means of the general
formula
\begin{equation}
\int d\Omega Y_{l}^{m\ast }\left( \theta ,\phi \right)
Y_{l_{1}}^{m_{1}}\left( \theta ,\phi \right) Y_{l_{2}}^{m_{2}}\left( \theta
,\phi \right) =\sqrt{\frac{\left( 2l_{1}+1\right) \left( 2l_{2}+1\right) }{%
4\pi (2l+1)}}%
C_{l_{1}l_{2};00}^{l_{1}l_{2};l0}C_{l_{1}l_{2};m_{1}m_{2}}^{l_{1}l_{2};\;l%
\;m},  \label{g6}
\end{equation}

\bigskip

(where $C$ symbols correspond to Clebsch-Gordan coefficients) and of
the property $Y_{l}^{-m}\left( \theta ,\phi \right) =\left( -1\right)
^{m}Y_{l}^{m\ast }\left( \theta ,\phi \right) $.

To put better in evidence the symmetries Clebsch-Gordan coefficients can be
rewritten in terms of $3j-$Wigner symbols\ :

\begin{equation}
C_{l_{1}l_{2};m_{1}m_{2}}^{l_{1}l_{2};\;l\;m}=\left( -1\right)
^{l_{1}-l_{2}+m}\sqrt{2l+1}\left(
\begin{array}{ccc}
l_{1} & l_{2} & l \\
m_{1} & m_{2} & -m%
\end{array}%
\right) .  \label{g7}
\end{equation}

\bigskip

By collecting all these properties and substituting in Eq. (\ref{g5}) we
obtain for the matrix element (\ref{g2}):

\begin{eqnarray}
&&\sum_{l=0,1}\sqrt{\left( 2l_{i}+1\right) \left( 2l_{j}+1\right) \left(
2l_{i^{\prime }}+1\right) \left( 2l_{j^{\prime }}+1\right) }\times  \notag \\
&\times &\left(
\begin{array}{ccc}
l_{j} & l_{j^{\prime }} & l \\
0 & 0 & 0%
\end{array}%
\right) \left(
\begin{array}{ccc}
l_{i} & l_{i^{\prime }} & l \\
0 & 0 & 0%
\end{array}%
\right) \underbrace{(-1)^{2(l_{i}-l_{i^{\prime }})}}_{=1}\underbrace{%
(-1)^{2(l_{j}-l_{j^{\prime }})}}_{=1}\sum_{m=-l}^{l}(-1)^{-m-m_{i}-m_{j}}%
\left(
\begin{array}{ccc}
l_{i} & l_{i^{\prime }} & l \\
-m_{i} & m_{i^{\prime }} & -m%
\end{array}%
\right) \left(
\begin{array}{ccc}
l_{j} & l_{j^{\prime }} & l \\
-m_{j} & m_{j^{\prime }} & m%
\end{array}%
\right) \times  \notag  \label{g8} \\
&\times &\int_{0}^{\infty }dr_{1}\left[ \frac{1}{r_{1}^{l-1}}%
\int_{0}^{r_{1}}dr_{2}r_{2}^{l+2}R_{j}^{\ast }\left( r_{2}\right)
R_{j^{\prime }}\left( r_{2}\right) +r_{1}^{l+2}\int_{r_{1}}^{\infty }dr_{2}%
\frac{1}{r_{2}^{l-1}}R_{j}^{\ast }\left( r_{2}\right) R_{j^{\prime }}\left(
r_{2}\right) \right] R_{i}^{\ast }\left( r_{1}\right) R_{i^{\prime }}\left(
r_{1}\right) .
\end{eqnarray}

\bigskip

Here the sum over $l$ has been restricted to the first $2$ terms by using
the triangular property of $3j$-Wigner symbols:

\begin{eqnarray*}
\left\vert l_{i^{\prime }}-l\right\vert &\leq &l_{i}\leq l+l_{i^{\prime }},
\\
\left\vert l_{j^{\prime }}-l\right\vert &\leq &l_{j}\leq l+l_{j^{\prime }}.
\end{eqnarray*}

\bigskip

Finally, the last integrals can be evaluated numerically.



\end{document}